# Constructing Sub-scale Surrogate Model for Proppant Settling in Inclined Fractures from Simulation Data with Multi-fidelity Neural Network

Pengfei Tang[1], Junsheng Zeng[2*], Dongxiao Zhang[3*], and Heng Li[4]

[1] Department of Energy and Resources Engineering, College of Engineering, Peking University, Beijing 100871, P. R. China

[2] Intelligent Energy Laboratory, Peng Cheng Laboratory, Shenzhen 518000, P. R. China

[3] School of Environmental Science and Engineering, Southern University of Science and Technology, Shenzhen 518055, P. R. China

[4] School of Earth Resources, China University of Geosciences, Wuhan 730074, P. R. China

Corresponding authors:
Junsheng Zeng: zengjsh@pcl.ac.cn
Dongxiao Zhang: zhangdx@sustech.edu.cn

**Abstract**

Particle settling in inclined channels is an important phenomenon that occurs during hydraulic fracturing of shale gas production. Generally, in order to accurately simulate the large-scale (field-scale) proppant transport process, constructing a fast and accurate sub-scale proppant settling model, or surrogate model, becomes a critical issue, since mapping between physical parameters and proppant settling velocity is complex. Previously, particle settling has usually been investigated via high-fidelity experiments and meso-scale numerical simulations, both of which are time-consuming. In this work, a new method is proposed and utilized, i.e., the multi-fidelity neural network (MFNN), to construct a settling surrogate model, which could utilize both high-fidelity and low-fidelity (thus, less expensive) data. The results demonstrate that constructing the settling surrogate with the MFNN can reduce the need for high-fidelity data and thus computational cost by 80%, while the accuracy lost is less than 5% compared to a high-fidelity surrogate. Moreover, the investigated particle settling surrogate is applied in macro-scale proppant transport simulation, which shows that the settling model is significant to proppant transport and yields accurate results. This opens novel pathways for rapidly predicting proppant settling velocity in reservoir applications.

## 1. Introduction

The Boycott effect, which describes settling behavior of particles in inclined containers, is a vital phenomenon in engineering, such as waste water treatment (Acrivos and Herbolzheimer, 1979; Sarkar et al., 2007), fuel cells (Wang et al., 2018), coal ash treatment (Ni et al., 2019), and proppant transport in hydraulic fracturing (Ba Geri et al., 2018; Chun et al., 2020; Kou et al., 2018). Particularly for proppant transport, the settling process is constrained in extremely narrow channels (fractures). Fracture inclination leads to complex granular-induced instabilities and undesired acceleration of settling at large-scale (Zeng et al., 2021b). Despite its significance, the Boycott effect has been largely ignored in previous works when simulating proppant transport at large-scale. To the best of the authors' knowledge, the settling model of proppant transport in inclined fractures has not yet been systematically investigated and determined in extant literature. Therefore, it is necessary to construct a new settling model which considers this effect for more accurate macro-scale proppant transport modeling.

Essentially, two genres of approaches exist to investigate the proppant settling model in inclined fractures, i.e., physical experiments and numerical simulations. The experimental approach is a direct and accurate method to obtain intuitive and reliable results (Ba Geri et al., 2018; Chun et al., 2020; Kou et al., 2018). However, physical experiments are commonly constrained by numerous factors, including length scale, time cost, and experimental limits. In particular, frequently changing physical properties of carrier/proppant and fulfilling specified boundary and initial conditions constitute markedly difficult tasks. Moreover, accurate velocity measurement presents another critical issue when a large number of particles exist. In addition to the experimental approach, numerical simulation has been utilized as an efficient strategy to investigate proppant transport in recent decades. In most cases, refined simulations can accurately recover phenomena in physical experiments. Furthermore, the numerical approach is more flexible in designing various physical properties and injection plans. It is also convenient for obtaining more details for elucidating complex processes with numerical simulations, such as sand bed packing and erosion. Especially for studying the proppant settling process in inclined fractures, the CFD-DEM method is proven to be an efficient numerical approach for capturing particle clustering behaviors (Zeng et al., 2016; Zeng et al., 2019; Zeng et al., 2021a), which is adopted to obtain raw data for constructing settling model in this work.

In order to construct the model of proppant settling in inclined fractures, many factors should be considered, such as Archimedes number *Ar*, Galilei number *Ga*, size ratio *Sr*, inclination angle $\theta$, etc. In other words, the dimension of free parameter space is high, and obtaining a full-parameter settling model requires sufficient valid supporting data from simulations in the parameter space. However, the computational and time costs of numerical simulations would be high when a large number of simulations with a refined mesh are required to construct the settling model. To solve this issue, the parameter orthogonal hypothesis is usually adopted to first convert this problem into one-dimensional fitting for various parameters, and then the multi-variable settling model is obtained by simple algebraic multiplication. This strategy however, is too idealized, and is not applicable for most complex high-dimensional mappings.

To overcome this challenge, recently, fast data-driven surrogate models are widely applied in investigations of particle-laden flows (Das et al., 2018; Lu et al., 2012; Sen et al., 2018a; Sen et al., 2018b). In these works, the multi-fidelity (MF) concept is introduced to greatly save computational time and resources without loss of accuracy for the obtained models. In the multi-fidelity surrogate model, there are two sub-models, called the low-fidelity (LF) sub-model and the high-fidelity (HF) sub-model. First, the LF model is developed with inexpensive data from 3D coarse grid simulations or 2D simplified simulations. Although LF data are less accurate, the evolution trend of LF data approximately coincides with that of HF data. Then, HF data from fine grid numerical simulations are fed into the HF model and used to rectify the error of the LF model. For constructing LF/HF sub-models, various conventional regression/fitting approaches, such as modified Bayesian Kriging (Das et al., 2018; Lu et al., 2012; Sen et al., 2018a; Sen et al., 2018b) and neural network (Meng and Karniadakis, 2020), can be adopted based on the data amount.

In this paper, we aim to construct a fast surrogate model for describing the settling behavior of proppant in inclined fractures based on the multi-fidelity neural network (MFNN) framework (Meng and Karniadakis, 2020), considering that with sufficient data, the neural network is highly capable of representing complex non-linear mappings. First, HF/LF data are obtained from 3D fine-mesh/coarse-mesh unresolved CFD-DEM simulations. Second, the single-parameter surrogate model is constructed for fine-tuning hyper-parameters in MFNN, as well as for validation of the presented framework. Third, a bi-parameter surrogate model is obtained for practical applications.

The remainder of this work proceeds as follows: in section 2, the problem setting, the unresolved CFD-DEM method, and the MFNN framework are introduced. In section 3.1, qualitative and quantitative discussions are presented by comparing the HF and LF simulation results. In section 3.2, the construction process of the single-parameter (settling velocity vs. inclination angle) based on MFNN is demonstrated. In section 3.3, a bi-parameter (settling velocity vs. inclination angle and density ratio) surrogate model is constructed. In section 3.4, the settling model is utilized in a case study of field-scale proppant transport. Finally, we conclude the paper and provide directions for future research in section 4.

## 2. Methodology

### 2.1. Macroscopic simulation for proppant transport

First, we provide a brief overview of a conventional macroscopic simulation method, i.e., the concentration transport method, for field-scale proppant transport, and demonstrate how the sub-scale proppant settling model can be embedded in the large-scale simulation framework.

The concentration transport method is a typical large-scale method, in which the amount of proppant is represented by concentration in a volume-average manner. For simplicity, in this work, we consider a non-propagating fracture with constant fracture width. Then, the motion for slurry, i.e., the mixture of carrier fluid and proppant, is governed by two-dimensional mass-conservation and quasi-Darcy equations:

$$\nabla \cdot \mathbf{U}_{sl} = q_{inj}, \tag{1}$$

$$\mathbf{U}_{sl} = -\frac{W^2}{12\mu_{sl}}\nabla(P - \rho_{sl}gh), \tag{2}$$

where $\mathbf{U}_{sl} = C\mathbf{U}_p + (1-C)\mathbf{U}_f$ is the 2D slurry velocity, in which $C$ indicates the proppant concentration; $\mathbf{U}_p$ and $\mathbf{U}_f$ are the average fluid and proppant velocity, respectively; $\mu_{sl} = \mu_f \Big/ \left(1 - \frac{C}{C^*}\right)^{1.82}$ is the effective viscosity, in which $\mu_f$ is the fluid dynamic viscosity and $C^*$ is the proppant packing limit; $\rho_{sl} = C\rho_p + (1-C)\rho_f$ is the slurry density; $q_{inj}$ is the source term due to proppant/carrier injection; $W$ is the width of the fracture; $g$ is the magnitude of gravity acceleration; and $h$ is the scalar vertical height.

For the particle phase, proppant concentration is also governed by the mass balance equation:

$$\frac{\partial C}{\partial t} + \nabla \cdot (C\mathbf{U}_p) = Cq_{inj}. \tag{3}$$

Usually, proppant velocity is assumed to be directly related to the velocity of slurry. Particularly, proppant velocity is considered the same as slurry in the horizontal direction. In the vertical direction, however, the velocity of proppant is considered to be the combination of slurry vertical velocity and proppant settling velocity:

$$\mathbf{U}_p = \mathbf{U}_s + w_s \vec{z}, \tag{4}$$

where $w_s$ is the proppant settling velocity.

The governing equation of proppant concentration degenerates into a passive transport equation by substituting equation (4) into equation (3). Specifically, once the slurry velocity is obtained by solving equations (1) and (2), the proppant concentration can be explicitly updated by solving equation (3). However, in order to solve equation (3), it is necessary to first calculate the proppant settling velocity. In general, proppant settling velocity is affected by numerous factors, including proppant and carrier fluid properties, fracture roughness and inclination, etc. In this work, we mainly consider the fracture inclination effect on proppant settling. The primary data describing particle settling for various parameters are firstly obtained through refined meso-scale simulation methods, and we aim to construct a fast and accurate proppant settling surrogate for a larger parameter range based on the raw data.

**2.2. Mesoscopic simulation for proppant transport**

As illustrated in Figure 1, the concerned meso-scale proppant settling process is constrained in a periodic inclined narrow fracture. Here, the intersection angle between the fracture surface and gravity direction is defined as $\frac{\pi}{2} - \theta$, where $\theta$ is the inclination angle. Then, $\theta = 90°$ indicates a rigorously vertical state. Initially, proppant particles are evenly distributed in the domain, which can be considered as a suspension. Then, due to gravity, proppant will settle in the $y$- and $z$-directions. For this simulation domain, two sides in the $y$-direction are set as non-slip solid boundaries, and the other four sides are set as periodic boundaries both in the $x$- and $z$-directions.

It should be noted that the characteristic length scale of the unresolved CFD-DEM method is usually limited to several decimeters because of the meso-scale constraints. If the two sides in the settling direction ($z$-direction) are set as solid boundaries, proppant will settle on the domain bottom very rapidly. The time period is too short for the development of granular-induced instabilities, leading to an inaccurate average settling velocity, which differs from the actual results in field-scale. Periodic boundaries in the settling direction allow both fluid and particles to continuously migrate through the domain without solid barriers. After the two-phase system reaches a quasi-steady state, the credible statistical-average quantities can then be easily calculated.

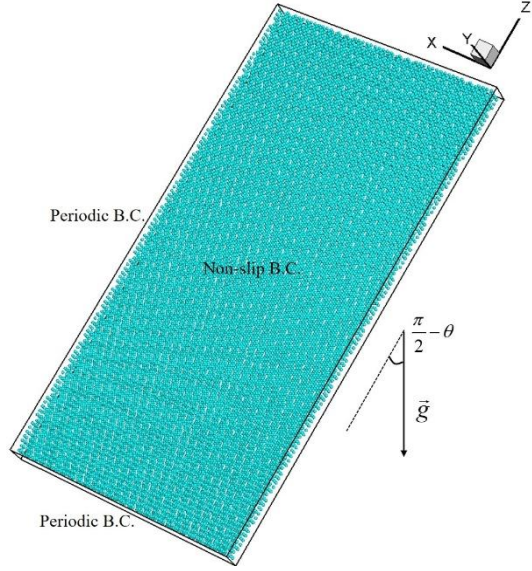

**Figure 1.** Illustration of domain settings for proppant settling in inclined periodic channels. $\vec{g}$ is the gravity acceleration.

As investigated by previous work (Zeng et al., 2021b), further granular-induced Rayleigh-Taylor and Kelvin-Helmholtz instabilities can develop in this process, which leads to particle-clustering behaviors and greatly changes the settling pattern of proppant in inclined fractures. It is critical to construct a data-driven surrogate to describe this novel settling law from simulation data. Without loss of generality, through non-dimension analysis, the dimensionless numbers dominated in this process, except for inclination angle $\theta$ and particle volume fraction $\alpha_p$, are listed as follows:

$$Ga = \frac{\sqrt{gW^3}}{\nu_f}, \quad D_r = \frac{\rho_p}{\rho_f}, \quad S_r = \frac{D}{W}, \tag{5}$$

where $Ga$ is the Galilei number; $D_r$ is the ratio of density; $S_r$ is the ratio of size; $D$ is the diameter of particle; $W$ is the fracture width; $\rho_p$ and $\rho_f$ are the density of particle and fluid density, respectively; and $\nu_f$ is the kinetic viscosity of fluid.

In this paper, we are mainly concerned about the influence of inclination angles and density ratios on proppant settling in inclined fractures. Our aim is to construct a data-driven surrogate for the acceleration ratio, which is defined as the ratio of average settling velocity in inclined cases over that in vertical cases:

$$Acc_r = \frac{w_s}{w_s^{\theta=\pi/2}} = f(\theta, D_r), \tag{6}$$

where $Acc_r$ is the settling speed-up ratio; $w_s$ is the average proppant settling velocity in inclined cases; and $w_s^{\theta=\pi/2}$ is the average proppant settling velocity in vertical cases.

In order to calculate the proppant settling velocity in inclined fractures mentioned in equation (6), here we apply a refined Eulerian-Lagrangian method to solve the issue, i.e., the unresolved CFD-DEM method. Here, the term "unresolved" indicates that the fluid field around the proppant surface is not resolved under the CFD-DEM framework. Instead, fluid-particle interaction forces are modeled as drag force, lift force, pressure gradient force, etc. Different from the aforementioned macro-scale approach, the unresolved CFD-DEM method is a typical meso-scale numerical approach (Blais et al., 2016; Kuruneru et al., 2019; Vångö et al., 2018). On the one hand, the unresolved CFD-DEM is sufficiently accurate to describe the dynamic behaviors of proppant particles compared to macro-scale approaches, such as the two-fluid model (TFM) and the multi-phase particle-in-cell (MP-PIC) method. On the other hand, compared to micro-scale numerical approaches, such as particle-resolved direct numerical simulation (PR-DNS), it is more computationally-efficient, and is capable of simulating a large number of particles which can reach to millions. Therefore, the unresolved CFD-DEM method constitutes an ideal approach for investigating granular-induced instabilities and particle-clustering phenomena in inclined fractures.

In the unresolved CFD-DEM method, fluid motion is governed by a volume-average Navier-Stokes equation (VANS), as follows:

$$\frac{\partial \alpha_f}{\partial t} + \nabla \cdot \left( \alpha_f \mathbf{u}_f \right) = 0, \tag{7}$$

$$\frac{\partial \left( \alpha_f \mathbf{u}_f \right)}{\partial t} + \nabla \cdot \left( \alpha_f \mathbf{u}_f \mathbf{u}_f \right) = -\frac{\alpha_f}{\rho_f} \nabla p_f + \alpha_f \nabla \cdot \mathbf{T} + \alpha_f \mathbf{g} + \mathbf{f}_p, \tag{8}$$

where $\alpha_f$ is the fluid volume fraction; $\mathbf{u}_f$ is the velocity of fluid; $\rho_f$ and $p_f$ are the density and pressure of fluid, respectively; $\mathbf{g}$ is the gravity acceleration; $\mathbf{f}_p$ is the fluid-particle interaction force; and $\mathbf{T} = \nu_f \left( \nabla \mathbf{u}_f + \nabla \mathbf{u}_f^T \right)$ is the viscous tensor of incompressible Newtonian fluid.

For the particle phase, Newton's 2nd law is adopted to describe particle motion:

$$m_p \frac{d\mathbf{u}_p}{dt} = \mathbf{F}_H + \sum \mathbf{F}_C + \mathbf{g}, \tag{9}$$

$$\frac{d\mathbf{x}_p}{dt} = \mathbf{u}_p, \tag{10}$$

$$I_p \frac{d\boldsymbol{\omega}_p}{dt} = \sum \mathbf{T}_C + \mathbf{T}_{RF}, \tag{11}$$

where $m_p$ is the mass of proppant particle; $\mathbf{u}_p$ is the velocity of particle; $\mathbf{F}_H$ is the hydrodynamic force due to fluid-particle interaction; $\mathbf{F}_C$ is the collision force due to particle-particle or particle-wall interaction; $\mathbf{x}_p$ is the particle displacement; $I_p$ is the particle's moment of inertia; $\mathbf{\omega}_p$ is the angular velocity of particle; and $\mathbf{T}_C$ and $\mathbf{T}_{RF}$ are the collision torque and rolling friction torque, respectively.

In this paper, the soft-sphere model based on the discrete element method (DEM) is adopted to calculate the repulsive and damping forces during particle-particle-wall collisions. For example, contact force in the normal direction of particle $i$ and particle $j$ can be calculated by:

$$\mathbf{F}_{Cn,ij} = k_n \mathbf{\delta}_n + \gamma_n \mathbf{u}_{ij}, \tag{12}$$

where the subscript '$n$' indicates the normal direction; $k_n$ and $\gamma_n$ are the stiffness and damping coefficients, respectively; $\mathbf{\delta}_n$ is the normal overlap between particle $i$ and particle $j$; and $\mathbf{u}_{ij}$ is the relative velocity of two particles $i$ and $j$.

Because the hydrodynamics forces due to fluid-particle interactions are not directly resolved in the unresolved CFD-DEM method, it is necessary to introduce appropriate force models to capture the fluid-particle coupling behaviors. Usually, only pressure gradient force and drag force are considered as the most important coupling mechanisms in previous works for simulating two-phase flows. However, considering the facts that the shear effect of fluid motion is severe and the density ratio is relatively small ($\rho_p/\rho_f < 5$) for proppant motion in inclined channels, the Saffman lift force and the added mass force should also be considered. Therefore, in this work, the hydrodynamic forces exerted on particles are modelled as a combination of Di Felice drag force, pressure gradient force, viscous tensor gradient force, Saffman lift force, and added mass force. Modelling details of these forces are not the main focus of this work, and interested readers may refer to Zeng et al. (2016) for additional details.

The proppant settling velocity mentioned in equation (4) can be calculated through mesoscopic simulation. The simulations with fine grid and small time-step can produce accurate and high-quality results, which is termed high-fidelity (HF) data. However, HF data usually incur substantial computational time and resources. Meanwhile, simulations with coarse grid and large time-step would lead to less accurate results, which is termed low-fidelity (LF) data. LF data are less expensive and easier to obtain compared to HF data. Consequently, we propose a method to make use of both expensive HF data and inexpensive LF data.

**2.3. Multi-fidelity Neural Network (MFNN)**

In this paper, we utilize the MFNN framework, which can combine LF data and HF data, to construct the particle settling surrogate model. Both the HF data and the LF data could be obtained through the numerical simulations introduced in section 2.2. Although a similarity or relationship usually exists between LF data and HF data, such a relationship is complicated and difficult to describe. Multi-fidelity modeling techniques are used to approximate this relationship (Das et al., 2018; Meng and Karniadakis, 2020). In multi-fidelity modeling techniques, a widely-used relationship between them can be expressed as follows:

$$y_H = \omega(x) y_L + b(x) \tag{13}$$

where $y_H$ and $y_L$ represents the HF and LF data, respectively; $\omega(x)$ is the operator, which approximates the correlation of multiplication; and $b(x)$ is the operator, which approximates the correlation of addition. The combination of multiplication and addition operator can approximate any linear function. However, there are many non-linear correlations in practice as a result of complex physical mechanisms. A more general expression for the correlation between HF data and LF data is given as:

$$y_H = M\ (x, y_L) \tag{14}$$

where $M$ is an operator, which involves both the linear and non-linear mapping from LF data to HF data. The operator $M$ can also be decomposed into the linear operator and nonlinear operator, which is shown as:

$$M = M_l + M_{nl} \tag{15}$$

In this paper, based on the above equations, MFNN combining HF data and LF data is proposed to construct surrogate settling models. The input data of MFNN are composed of the LF data and the HF data. The LF data are obtained from coarse grid simulations, while the HF data are obtained from fine grid simulations which are computationally-expensive. The architecture of the neural network is shown in Figure 2. The neural network that we adopted was MFNN, as shown in Figure 2, which was designed according to equation (15).

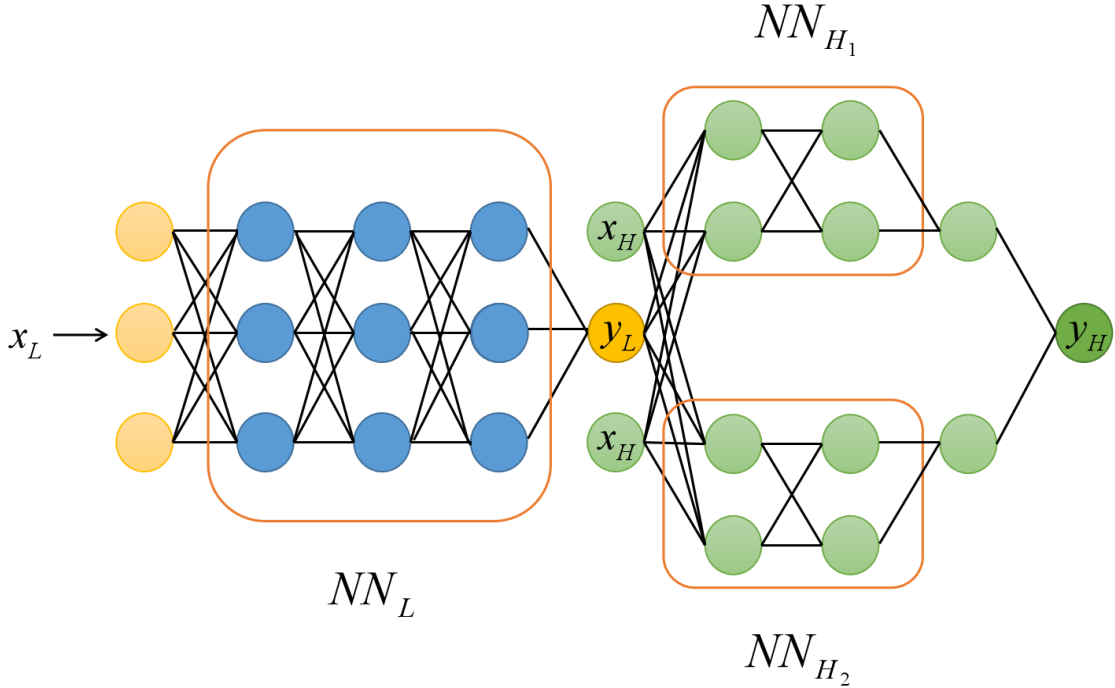

**Figure. 2.** Architecture of the MFNN.

The multi-fidelity network comprises three fully-connected neural networks. The first neural network $NN_L$ is trained to create a mapping between $x_L$ and $y_L$, which represent the input and the output of the LF data, respectively. The second neural network $NN_H$, composed of $NN_{H_1}$ and $NN_{H_2}$, is utilized to approximate the map from input to HF data. There is only one hidden layer with no activation function in $NN_{H_1}$. This is because $NN_{H_1}$ is designed to approximate the linear relationship between the LF and the HF data, while $NN_{H_2}$ is designed to approximate the nonlinear relationship between them. There are two hidden layers with 10 neurons in each layer. The combined output of $NN_{H_1}$ and $NN_{H_2}$ determine the final output $y_H$. Therefore, both the linear and non-linear correlation between the LF data and HF data can be approximated by the $NN_H$. When there are finite HF data, however, this would easily lead to overfitting in training the neural network with many parameters. As a consequence, in this paper, the training process is divided into two stages, i.e., the LF stage and the HF stage. In the LF stage, we train the parameters of $NN_L$ using a sufficient amount of LF data. After the LF stage is completed, the parameters of both $NN_{H_1}$ and $NN_{H_2}$ are trained using finite HF data while the parameters of $NN_L$ are fixed, which is called the HF stage. The design of the training stages could make the most use of the extant data to avoid overfitting. The LF neural network and the HF neural network are trained, respectively, by minimizing the loss functions, i.e., mean square error (MSE), which are defined as:

$$MSE_{y_L} = \frac{1}{N_{y_L}} \sum_{i=1}^{N_{y_L}} (|\hat{y}_L - y_L|^2) \qquad (16)$$

$$MSE_{y_H} = \frac{1}{N_{y_H}} \sum_{i=1}^{N_{y_L}} (|\hat{y}_H - y_H|^2) \qquad (17)$$

where $\hat{y}_L$ and $\hat{y}_H$ are the outputs of the $NN_L$ and $NN_H$, respectively. The loss function is optimized by the Adam optimizer. The activation function used in MFNN is a tangent function. The weights of both $NN_L$ and $NN_H$ are initialized with Xavier's initialization.

## 3. Results and Discussions
### 3.1. HF and LF unresolved CFD-DEM simulations

In this paper, we first use numerical simulations based on the unresolved CFD-DEM method to develop particle settling surrogate models. In the numerical simulations, in order to obtain more accurate simulation results, the continuous domain is usually divided with sufficiently-fine meshing. Continuous time is also divided into discrete small intervals. Then, the governing equations are solved on grids and discrete time steps. In general, the finer is the mesh and the smaller is the time step, the more accurate is the simulation result. However, substantial computational resources and time would be required to simulate a practical problem with fine mesh and small time-step, especially in the case of a high-dimensional problem. Indeed, the computational cost would increase exponentially with the increase of domain size and dimension. Therefore, it is expensive to attain accurate simulation results, which are called HF data. If we increase mesh size and time step, we would obtain simulation results with certain error, which are treated as LF data. Nevertheless, LF data, and especially trends of LF data, are also valuable for reference to practical problems. This is because the trends of LF data are highly similar to those of HF data. In this paper, we utilize 3D fine grid numerical simulation results as HF data. Compared to HF simulations, the size of the calculation region is reduced and the mesh is coarsened in LF simulations, while the volume fraction of particles remains unchanged. The detailed parameters and settings of HF simulations and LF simulations are shown in Table 1 and Table 2, respectively. Compared with HF numerical simulations, the volume of the calculation region in LF simulations is reduced by 10 times. Moreover, the size of the mesh is increased by two times in all three directions. In addition, the particles are packaged in all three directions. In other words, the original eight particles in HF simulations are represented as one particle in LF simulations. Therefore, the errors between HF and LF simulations are mainly introduced from the calculation area, mesh size, and particle size.

Table 1. Parameter settings for proppant settling in inclined fractures (high fidelity case).

| Parameter | Value | Parameter | Value |
| --- | --- | --- | --- |
| Fluid density | 1,000 kg/m³ | Particle density | 1,000 kg/m³~3,000 kg/m³ (interval: 200 kg/m³) |
| Fluid viscosity | 1 cp | Particle diameter | 0.6 mm |
| Domain size | $0.2 \times 0.1 \times 0.006$ m³ | Mesh size | $200 \times 100 \times 6$ |
| Stiffness coefficient | 3.5 N/m | Restitution coefficient | 0.1 |
| Friction coefficient | 0.6 | Rolling friction coefficient | 0.1 |
| Particle numbers | 120,000 | Simulation time | 20 s |
| CFD time step | 2 ms | DEM time step | 20 us |
| Inclination angles | 15°~90° (interval: 5°) | | |

Table 2. Parameter settings for proppant settling in inclined fractures (low fidelity case).

| Parameter | Value | Parameter | Value |
| --- | --- | --- | --- |
| Fluid density | 1,000 kg/m³ | Particle density | 1,000 kg/m³~3,000 kg/m³ (interval: 200 kg/m³) |
| Fluid viscosity | 1 cp | Particle diameter | 0.6 mm |
| Domain size | $0.06 \times 0.03 \times 0.006$ m³ | Mesh size | $30 \times 15 \times 3$ |
| Stiffness coefficient | 3.5 N/m | Restitution coefficient | 0.1 |
| Friction coefficient | 0.6 | Rolling friction coefficient | 0.1 |
| Particle numbers | 4,800 | Simulation time | 20 s |

| CFD time step | 2 ms | DEM time step | 20 us |
| Inclination angles | 15°~90° (interval: 5°) | | |

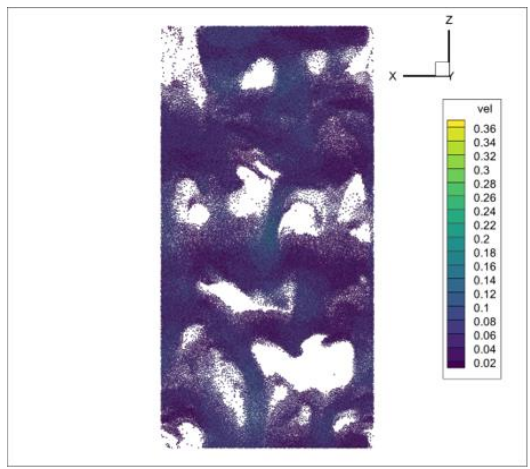

(a)

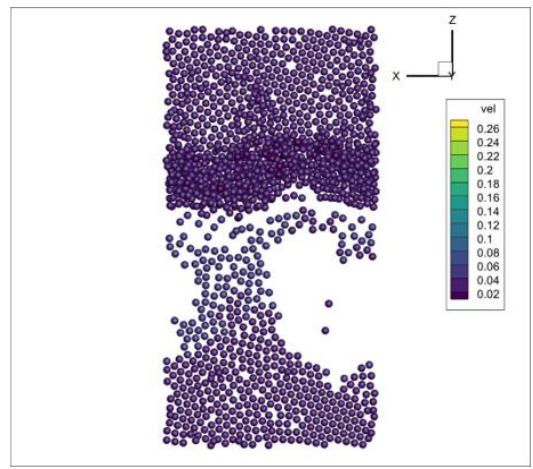

(b)

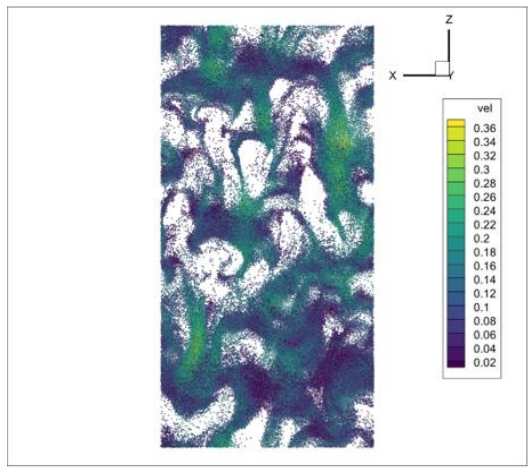

(c)

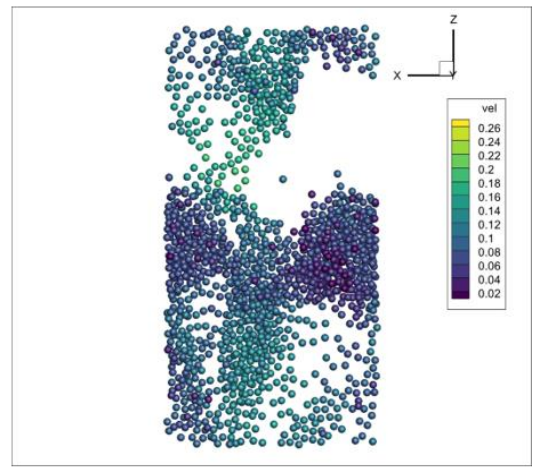

(d)

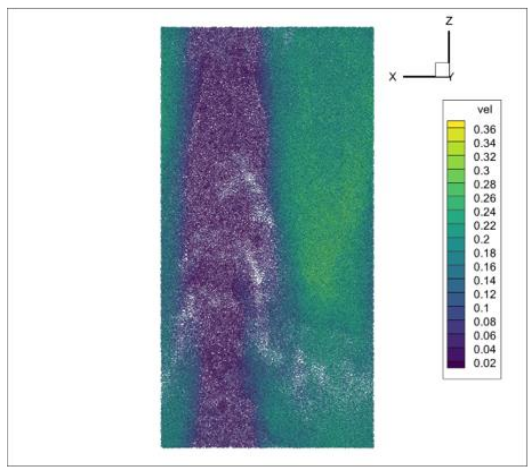

(e)

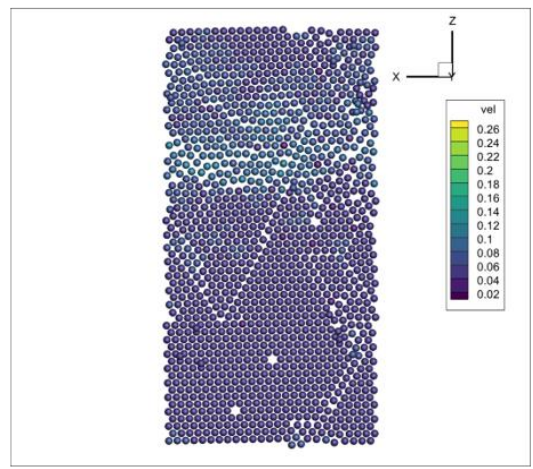

(f)

**Figure 3.** The proppant velocity distribution: (a, b) $\theta$=45° and $D_r$=1.2; (c, d) $\theta$=65° and $D_r$=1.6; (e, f) $\theta$=35° and $D_r$=2.6, of HF and LF unresolved CFD-DEM simulations, respectively.

Some simulation results of HF and LF are shown in Figure 3. We selected three different groups of results for demonstration. The left is the HF result, and the right is the LF result. Comparing the HF results with the LF results, it can be concluded that there are obvious differences between the two results, but certain similarities remain. The results of LF can still reflect the structure of particles to a certain extent, which can be regarded as the structure on a small scale. When the angle is large (e.g., 65°), LF data and HF data exhibit an obvious particle clustering phenomenon. When the fracture dip angle is small (e.g., 35°), the particles in LF and HF data spread on the fracture surface in the process of settling. The range of particle sedimentation velocity of HF and LF simulation with specific $\theta$ and $D_r$ is basically identical.

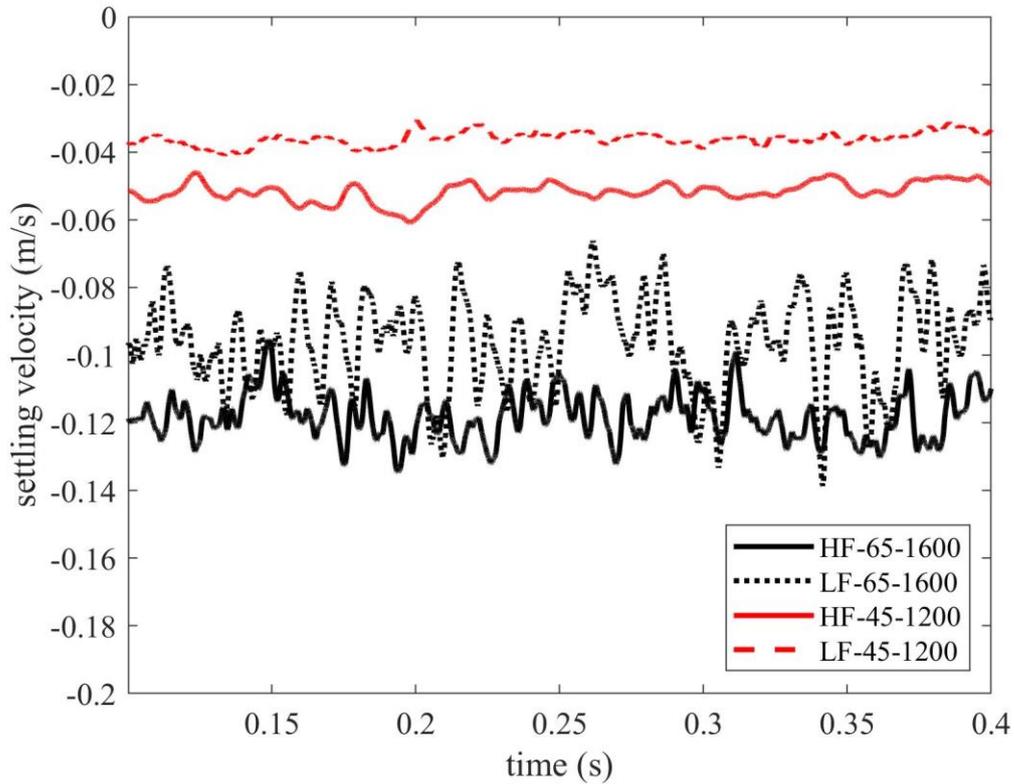

**Figure 4.** Average settling velocity of proppant in the case of HF and LF simulations.

In order to quantitatively describe and analyze the differences between HF data and LF data, we choose two groups of data, as shown in Figure 4. As shown in the figure, the setting velocity converges and oscillates around a certain number after 0.1 s for each simulation. In the case of $\theta$=65° and particle density=1600, the average particle settling velocity of LF and HF simulation converges and oscillates around -0.10 m/s and -0.12 m/s, respectively. In the case of $\theta$=45° and particle density=1200, the average particle settling velocity of LF and HF simulation converges and oscillates around -0.04 m/s and -0.05 m/s, respectively. "-" means the particle settling downward along the inclined fracture. The results show that both the magnitude and scope of the oscillations of settling velocity are similar. However, differences still exist between HF and LF data. The relative error of average settling velocity in these two cases can reach approximately 20%, which is too large to be neglected. Therefore, it is insufficient to use LF data directly for constructing an accurate particle settling model.

**3.2. Investigation of the surrogate settling model in 1D parameter space**

We first verify the validity of the model in one-dimensional parameter space. The accuracy and effectiveness of MFNN is demonstrated in approximating the correlation between the particle acceleration ratio and the inclination angle of fractures. We perform numerical simulations based on the unresolved CFD-DEM method to obtain the data. The coarse grid simulations and the fine gird simulations are carried out, respectively, with different angles. The particle acceleration ratio obtained from fine grid simulations can be treated as HF data, because the results of numerical simulations with fine grids are sufficiently accurate, while LF data from coarse gird simulations possess certain errors. For both HF and LF data, the inclination angle of the channel that we chose is from 15° to 90° with an interval of 5°. Specifically, 16 groups of data are as follows: [15, 20, 25, 30, 35, 40, 45, 50, 55, 60, 65, 70, 75, 80, 85, 90] is the inclination angle of the channel.

The results of 16 HF simulations are regarded as the reference shown in Figure 5, since they are fine enough to obtain accurate results, although they are computationally-intensive and time-intensive (Zeng et al., 2021b). The comparable cases include those from 16 groups of LF data only, three groups of HF data only, and a combination of three groups of HF data and 16 groups of LF data.

16 groups of LF data were adopted as the training set of $NN_L$ in order to construct the surrogate model of coarse gird simulation. The input $x_L$ = [15, 20, 25, 30, 35, 40, 45, 50, 55, 60, 65, 70, 75, 80, 85, 90] is the inclination angle of the channel. The output $y_L$ is the particle acceleration ratio. After the training of $NN_L$ is completed, the $y_L$ together with $x_H$ is passed to $NN_{H_1}$ and $NN_{H_2}$ as input. Three groups of HF data are utilized to train $NN_H$ to obtain an accurate prediction of acceleration ratio based on the inclination angle of the channel and the acceleration ratio with some errors. The input of the $NN_H$ is $x_H$ = [30, 50, 70].

First, we attempt to construct the settling surrogate model with the three groups of HF data only, which is shown in Figure 5 as HF. There are two hidden layers with 10 neurons in $NN_L$. Meanwhile, two hidden layers with 10 neurons and 5 neurons, respectively, are adopted in $NN_{H_2}$. There are two linear hidden layers with 10 neurons and 5 neurons in $NN_{H_1}$. There is no regularization of parameters in both $NN_L$ and $NN_H$. The learning rate is 0.001. As shown in Figure 5, the prediction of the model is accurate in the middle of the curve. However, when the angle is smaller than 30° or larger than 0.8, the prediction of the acceleration ratio is not accurate due to the lack of HF data. In addition, the mean relative square error of the model is 0.19. If only three groups of HF data are used, there will be a large error at the locations where there is no selected point, such as 15° and 90°, because the data are limited.

In order to obtain more accurate predictions, the LF data cooperating with the HF data are utilized in MFNN. As illustrated in Figure 5, accurate predictions can be obtained via MFNN. It is worth noting that both MF and HF in Figure 5 only use three groups of HF simulation results. When $x = 45°$, there is a difference between the output of the model and the reference. The main reason for this is that there is a sudden drop in LF data, which has a great influence on the multi-fidelity output. Moreover, the HF data chosen as training data do not include the point $x = 45°$. Consequently, the prediction here has not been corrected due to the lack of the data at this point. It is also worth noting that the mean relative square error of the multi-fidelity architecture is 0.06, which is much smaller than the network using three groups of HF data only or 16 groups of LF data only, i.e., HF and LF in Figure 5. Meanwhile, compared with the reference, MFNN could reduce 80% of the computational cost. This shows that MFNN combining HF data and LF data can be applied to this problem and significantly reduce the error.

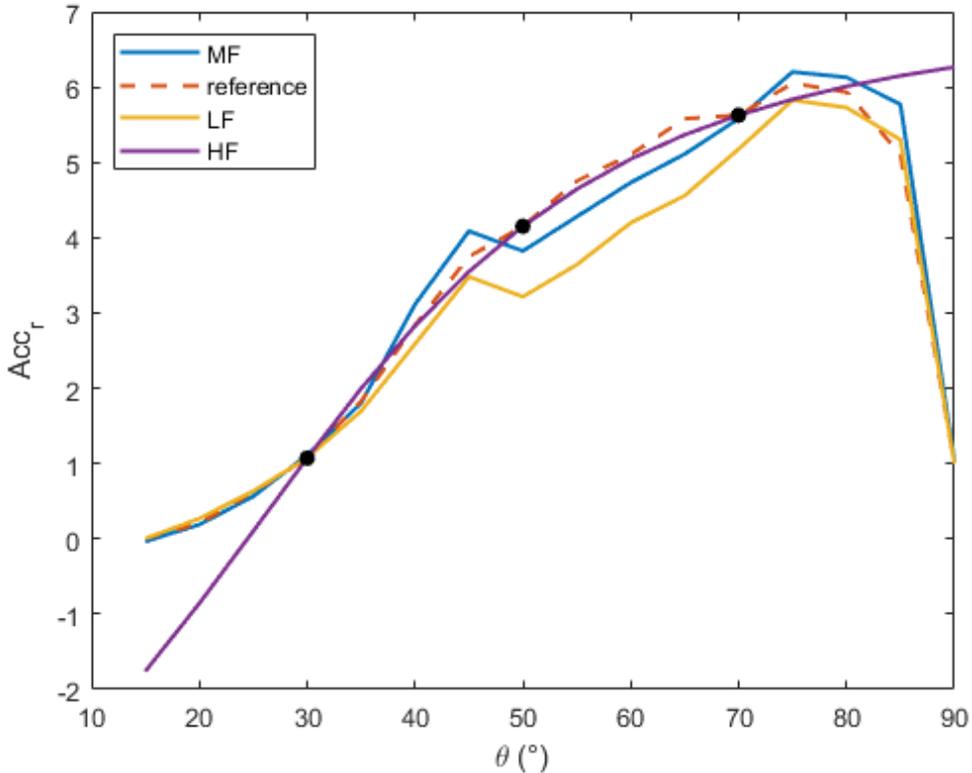

**Figure 5.** The one-dimensional settling surrogate model of fractures with different $\theta$ constructed via MFNN.

### 3.3. Investigation of the surrogate settling model in 2D parameter space

In this section, MFNN is extended to investigate the settling model in two-dimensional parameter space composed of inclination angle and particle density ratio. As mentioned previously, computational cost could be markedly high if many fine grid numerical simulations are performed to obtain accurate results when the parameter space is two-dimensional or higher.

Instead, we perform 160 groups of HF and LF numerical simulations based on the unresolved CFD-DEM method. The differences between HF and LF simulations include the difference of simulation region size, mesh size and particle size, while other parameters remain the same, e.g., the volume fraction of particles and other dimensionless number. In the training process, we first train the parameters of $NN_L$ with 80% of LF data and test the model with 20% of LF data. The input $x_1$ and $x_2$ represents the inclination angle of the channel and the density ratio of particles, respectively, while the output $y_L$ represents the predicted acceleration ratio by $NN_L$. We then train the parameters of $NN_H$ with 20% of HF data and test the MF model with 80% of HF data. The output of $NN_H$ is the final output of MFNN, denoted as $y_H$. It is worth noting that the LF and HF training data are obtained through random sampling. The detailed hyper-parameters of the training process are presented in Table 3. Here, we utilize Tanh and Adam as the active function and optimizer, respectively.

Table 3. Hyper-parameters of the training process.

| Hyper-parameters | Value of LF stage | Value of HF stage |
| --- | --- | --- |
| Batch size | 128 | 32 |
| Epoch | 10000 | 10000 |
| Learning rate | $1 \times 10^{-3}$ | $1 \times 10^{-3}$ |

The results of MFNN, LFNN, and the reference are shown in Figure 6. Here, we use the results of 160 HF simulations as the reference, i.e., the interval of $\theta$ and $D_r$ is 5° and 0.2, respectively, in the two-dimensional parameter space. Intuitively, the predicted acceleration ratio by MFNN is quite similar to the reference both in values and trends; whereas, the LF data are markedly different from the reference value, except for trends. The MSE of the output result of MFNN is 0.04, while the MSE of LF data is 0.30. It is worth noting that we use 32 groups of HF data and 128 groups of LF data to train MFNN. If we only use 32 groups of HF data to train the model, the MSE would be 0.09. In order to describe the results more accurately, we selected five curves at equal intervals and presented them in Figure 7. As shown in Figure 7, the overall MFNN results are very close to the reference, with only some local errors. This error is mainly due to the fact that there is no HF data sampling point in the vicinity of this region, and the true value is much higher than the known information (i.e., LF data).

To describe the results quantitatively, five groups of data are picked at equal intervals by particle density ratio, and plotted as Figure 7. The predicted acceleration ratio by MFNN is consistent with the reference, except when $D_r$ = 1.2 and angle = 55°~70°, which is resultant from HF data sampling, as shown in Figure 8.

The absolute error between MFNN output and the reference obtained with 160 groups of HF data, as well as the samples selected for training, are presented in Figure 8. The black points are the 32 groups of HF data used as training data, and the sampling method is random sampling. The sampling points are quite sparsely distributed in parameter space. Even so, the absolute error is relatively small in the region without sampling, except when $D_r$ = 1.2~1.6 and angle = 80°~90°. The acceleration ratio drops rapidly when $D_r$ = 1.2~1.6 and angle = 80°~90°, which makes prediction difficult. Moreover, there are no sampling data near this region, which contributes to the error.

From the perspective of relative error distribution, only the region in which angle = 15° has a relatively high relative error. This is because when angle = 15°, the acceleration ratio is close to 0, while the maximum value of the whole parameter space is 4.5. Therefore, a small absolute error near 0 will lead to a large relative error. Overall, the results of our model are accurate.

In the case mentioned above, the LF error comes from the calculation area size, mesh size and particle size, while other conditions, such as particle volume fraction, remain unchanged. We have successfully constructed the settling model via MFNN, which is proven to be both efficient and accurate. However, certain restrictions on LF and HF data still exist. The dimensionless numbers, except particle density ratio, of HF and LF simulations remain the same in the above investigation. As a consequence, the LF and HF data have a certain resemblance, which can be inferred from Figure 6. In order to further extend the application range of our model, compared to the previous section, the restrictions on LF data are relaxed here. In particular, the volume fraction of particle, which is an important factor in particle settling, is different in HF simulations and LF simulations.

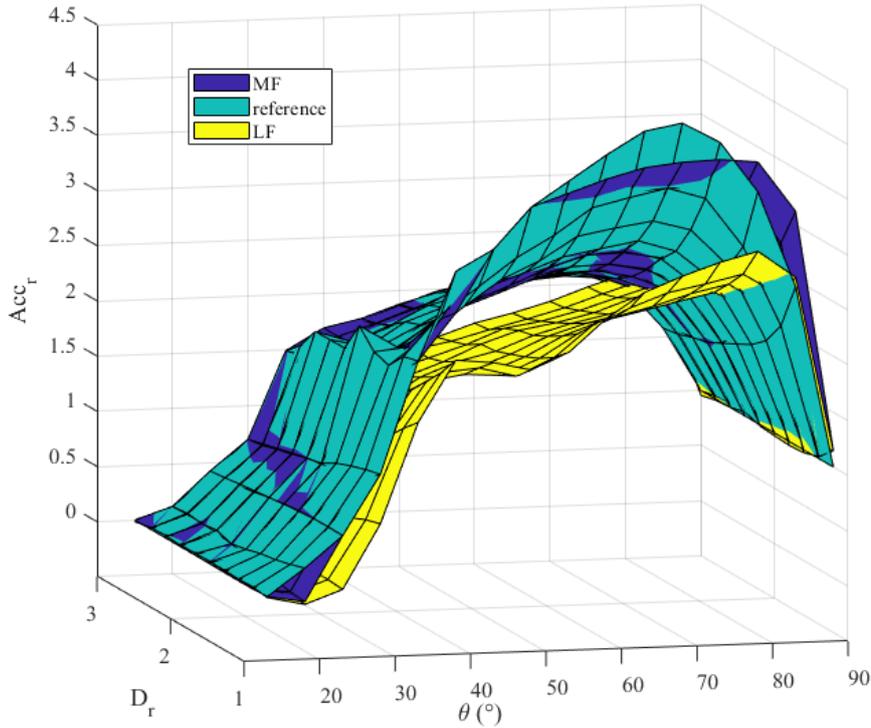

**Figure 6.** The acceleration ratio of particles investigated via LFNN and MFNN, as well as the reference obtained with 160 groups of HF data.

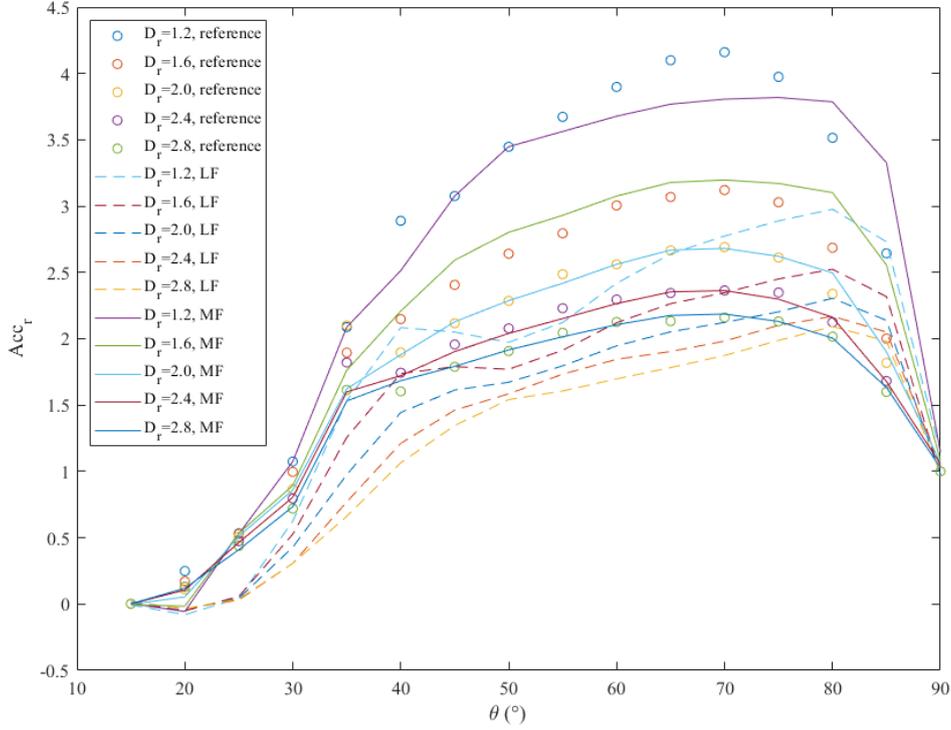

**Figure 7.** The acceleration ratio curve of particles investigated via LFNN and MFNN, as well as the reference obtained with 160 groups of HF simulations.

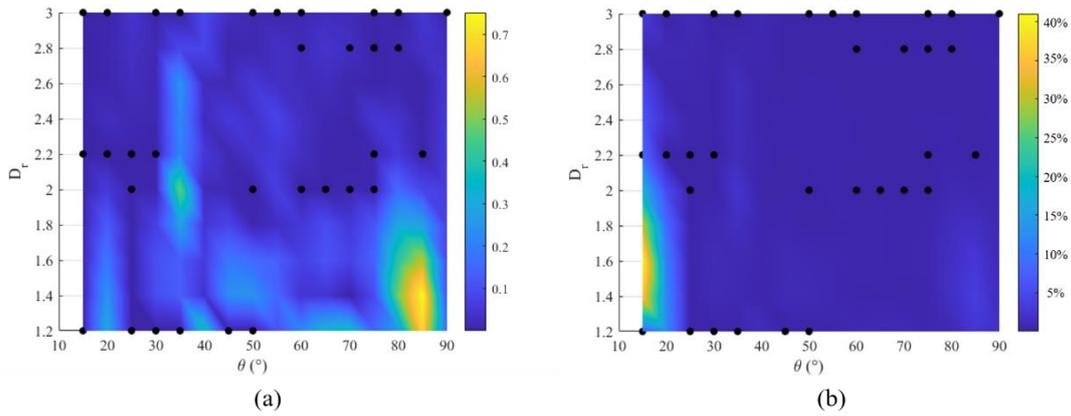

**Figure 8.** The absolute error of acceleration ratio (a) and the relative error of acceleration ratio (b) between MFNN output and the reference in two-dimensional space.

In order to further validate the effectiveness and robustness of MFNN, we design another case, in which for the same reference case, the particle volume fraction of the LF data differs from the reference, i.e., the error of the LF data is further increased. The MSE of the LF data increase from 0.30 to 0.41. The predicted acceleration ratio surface by MFNN and LFNN in parameter space is shown in Figure 9. There is a larger difference between the predicted acceleration ratio by LFNN and the reference. Nevertheless, the predicted acceleration ratio by MFNN is still consistent with the reference. From the error of the acceleration ratio curve shown in Figure 11(a), there is a certain error when $D_r$ = 1.2 and angle = 55°~70°, which is similar to Figure 8(a) and caused by the lack of sampling data. From the absolute error shown in Figure 11(a), there is another region with a certain error when angle = 30°~40°. That is probably because there is a lift of HF data in this region, while the LF data are gentle and without sudden lift. Moreover, there are no sampling data in this region. These two factors lead to the error when angle = 30°~40°. In the case of increasing error, the output of MFNN can still maintain similar accuracy.

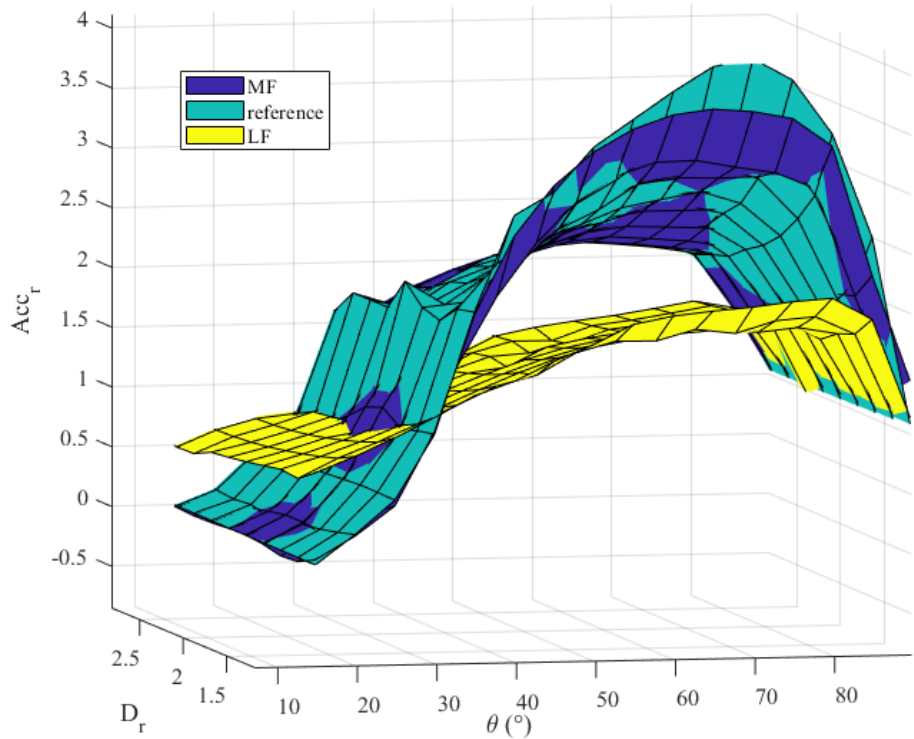

**Figure 9.** The acceleration ratio of particles investigated via LFNN and MFNN, as well as the reference obtained with 160 groups of HF data.

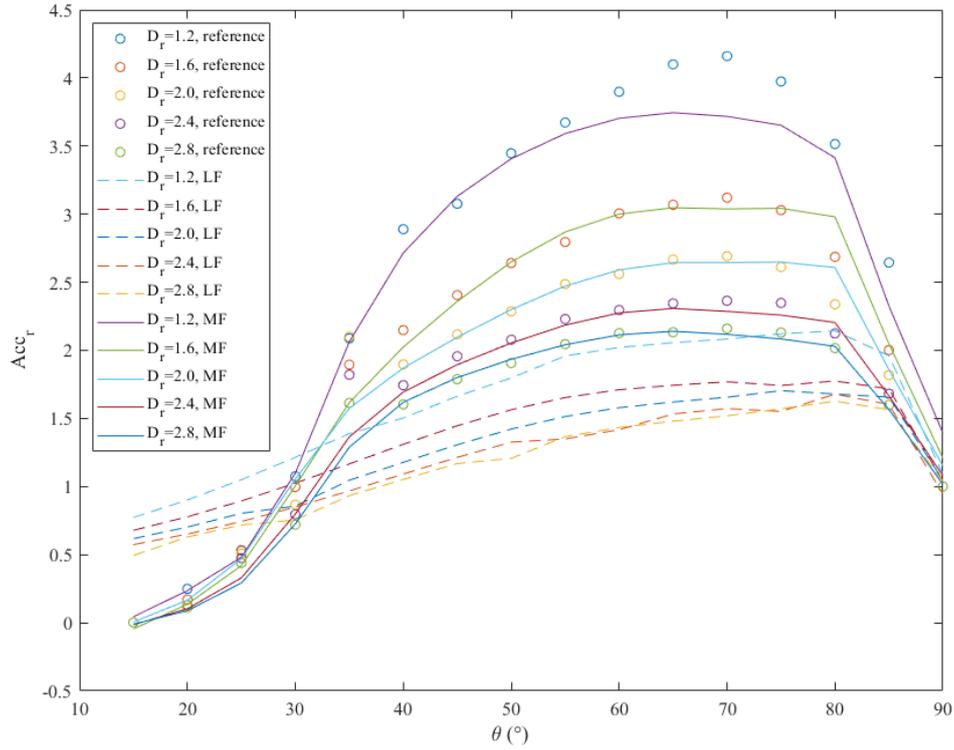

**Figure 10.** The acceleration ratio curve of particles investigated via LFNN and MFNN, as well as the reference obtained with 160 groups of HF simulations.

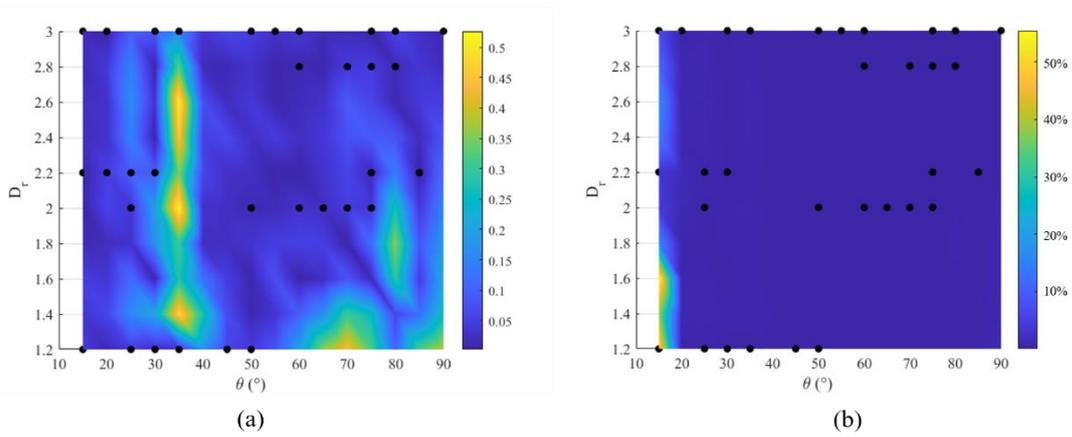

**Figure 11.** The absolute error of acceleration ratio (a) and the relative error of acceleration ratio (b) between MFNN output and the reference in two-dimensional space.

As indicated by the results of absolute error, the area with large absolute error is still the area not covered by the training sample. However, the regions with large relative errors are mainly distributed in regions where the acceleration ratio approaches 0. It can be seen that, although some additional error terms are introduced, our model can still extract the rules of LF data and apply them to HF data prediction.

## 3.4. Application of particle settling model in macro-scale proppant transport simulations

After constructing the particle settling model in inclined fractures using MFNN, we apply it to numerical simulations of macro-scale proppant transport to explore whether the particle settling model has any effect on macro-scale proppant transport. The process is as follows: first, we obtain the settling model, i.e., $Acc_r$, from large HF simulations, large LF simulations, and MFNN trained with small HF and large LF data, respectively. Subsequently, we apply the obtained settling model in macro-scale proppant transport simulations. Finally, we compare the results of different cases to draw conclusions. Here, the macro-scale simulations, which use 160 groups of HF simulation results as $Acc_r$, are treated as the reference, in which $Acc_r$ is obtained by HF meso-scale simulations only and then fed into macro-scale simulations. Similarly, the macro-scale simulations, which use 160 groups of LF simulation results as the parameter, are treated as the case based on $Acc_r$ via LF data.

Meanwhile, the case of MFNN is based on $Acc_r$ via MFNN trained using 32 groups of HF simulation results and 160 groups of LF simulation results. It is worth noting that, once MFNN is trained, the application of MFNN only costs several seconds to predict $Acc_r$. For comparison, we also consider cases in which the acceleration of particles is neglected, i.e., vertical cases. We selected two examples with different $D_r$ and angle for quantitative comparisons. In the first example, particles settle in the fracture with the angle of 45° and $D_r$ of 1.2, which is shown in Figure 12. For cases with lower density ratio, the settling velocity of particles is slower. It can be seen from the figures that the proppant distribution in the vertical fracture is significantly different from that in the inclined fracture, and the relative error of the sand bed height is 0.72. The results of MFNN are very close to the reference, and the relative error of sand bed height is only 0.02. However, the LF data are significantly different from the reference, with a relative error of 0.50. In the second case, particles settle in the fracture with an inclination angle of 65° and a $D_r$ of 1.6, which is shown in Figure 13. Because the particles are heavier, the settling rate is faster. When the particle sedimentation is faster, a similar transport phenomenon is observed. Meanwhile, the relative error between the reference and the vertical case is larger. The distribution of proppant in vertical fractures is significantly different from that in inclined fractures, and the relative error of sand bed accumulation height is 0.83. The results of MFNN are markedly close to the results of the reference, and the relative error of sand bed accumulation height is only 0.01. The relative error of the LF data is 0.54, which is significantly different from the reference.

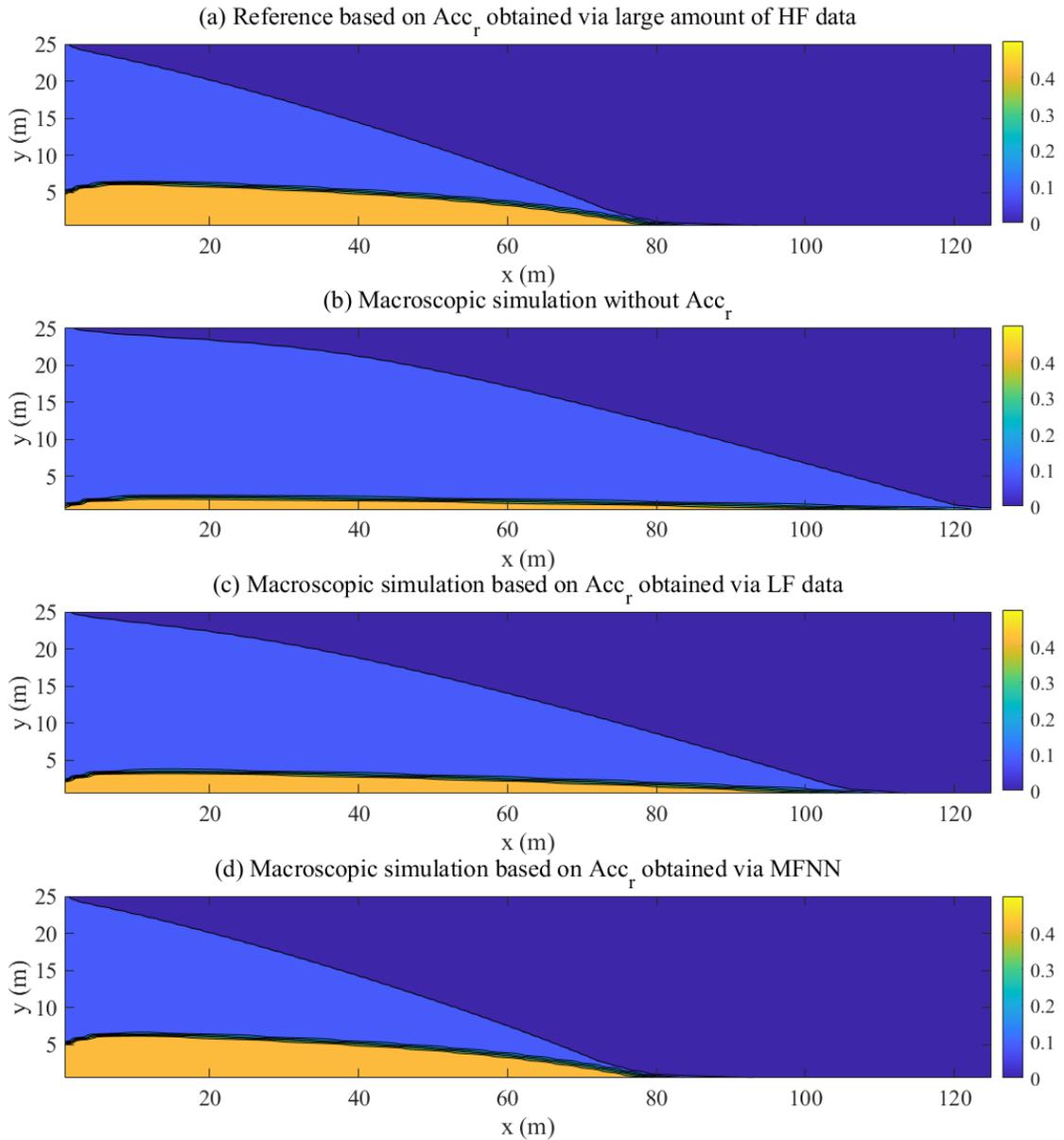

**Figure 12.** Comparison of particle volume fraction of macroscopic proppant simulations when $\theta$ = 45° and $D_r$ = 1.2: (a) reference based on $Acc_r$ obtained via a large amount of HF data; (b) without $Acc_r$; (c) based on $Acc_r$ obtained via LF data; (d) based on $Acc_r$ via MFNN.

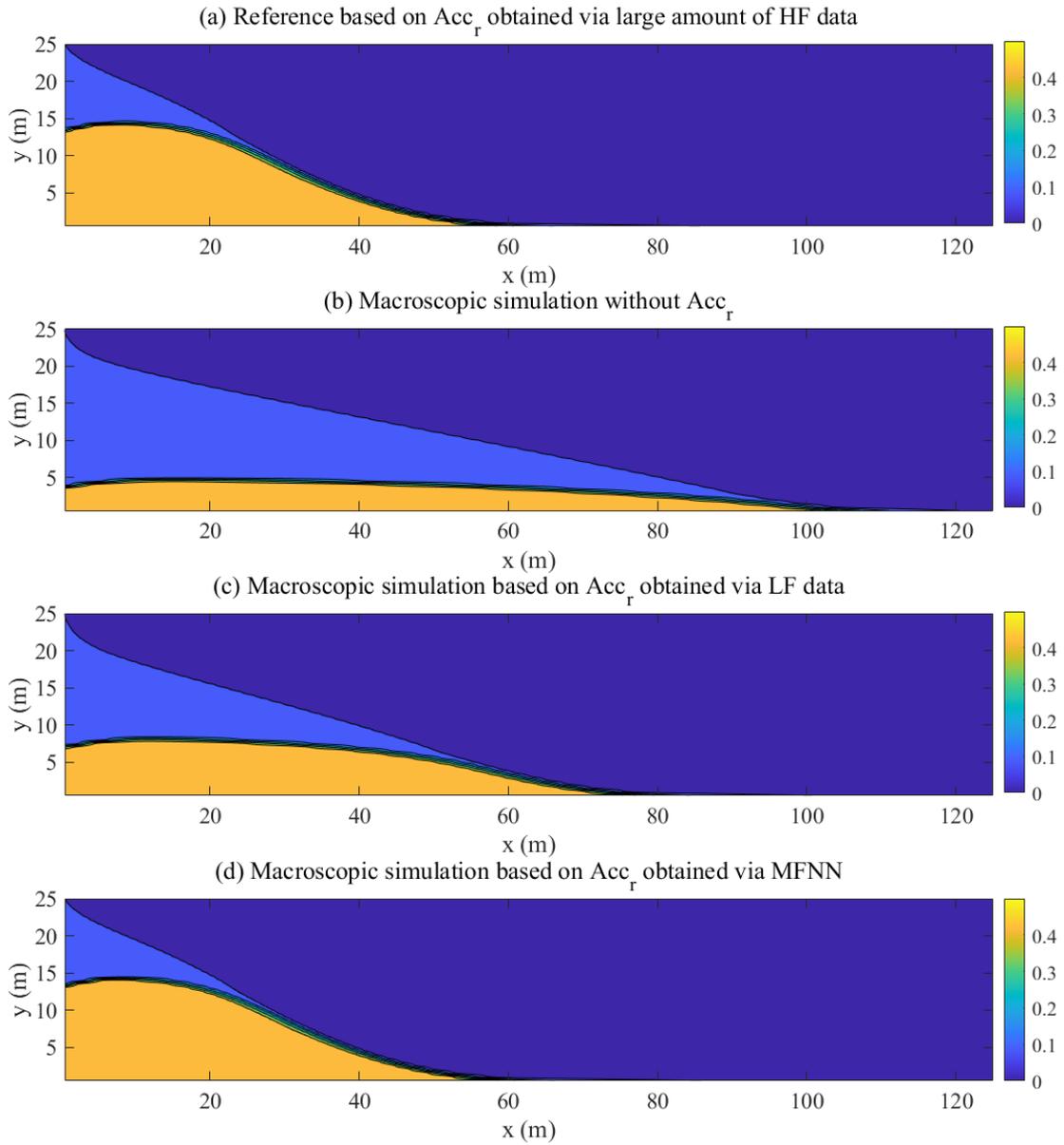

**Figure 13.** Comparison of particle volume fraction of macroscopic proppant simulations when $\theta$ = 65° and $D_r$ = 1.6: (a) reference based on $Acc_r$ obtained via a large amount of HF data; (b) without $Acc_r$; (c) based on $Acc_r$ obtained via LF data; (d) based on $Acc_r$ via MFNN.

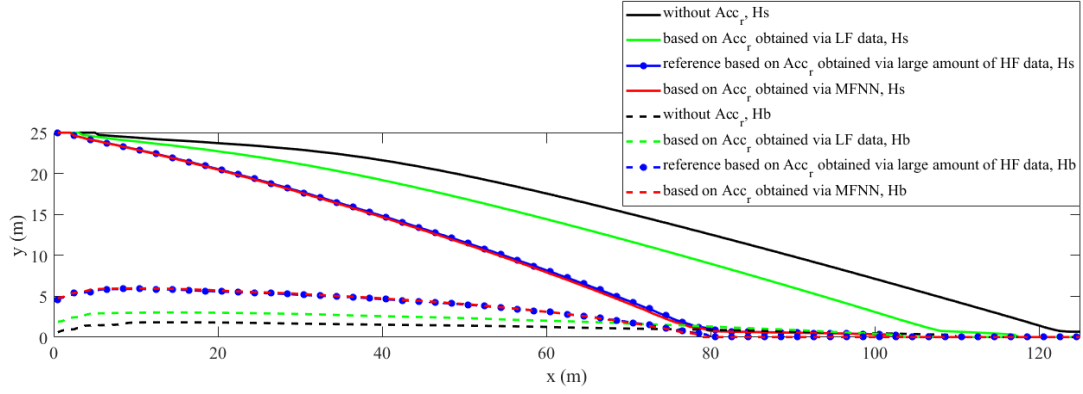

**Figure 14.** The height curve of sand bed (Hb) and slurry (Hs) in the case of vertical case, LF, MFNN, and the reference with $\theta = 45°$ and $D_r = 1.2$.

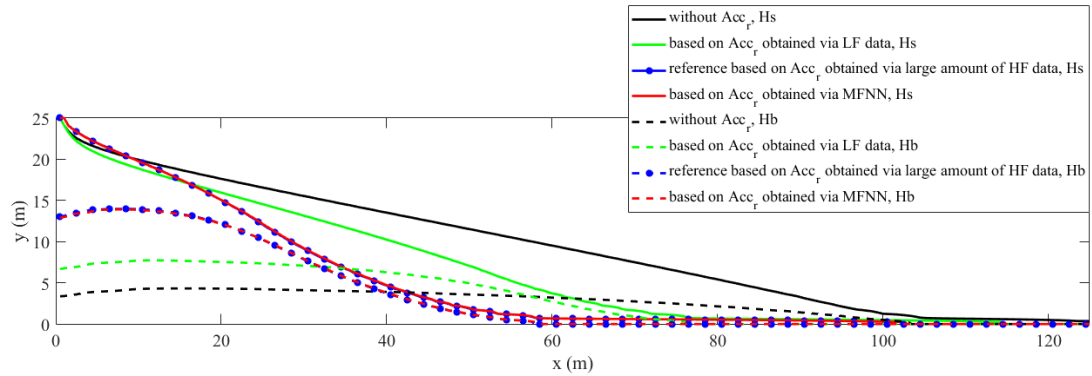

**Figure 15.** The height curve of sand bed (Hb) and slurry (Hs) in the case of vertical case, LF, MFNN, and the reference with $\theta = 65°$ and $D_r = 1.6$.

Figures 14 and 15 show the profiles of the sand bed accumulation height (Hb) and slurry front (Hs) of macro-scale proppant transport. The patterns and features of these two cases are different. In Figure 14, the particle density is close to the fluid, and the particle settling is much weaker. In contrast, the particle density is much larger than the fluid in Figure 15, which results in obvious sand bed packing. The solid line is the slurry front, and the dashed line is the sand bed accumulation profile. The red line is the results based on $Acc_r$ obtained via MFNN, and the blue line is the reference. These two lines basically coincide, which shows that MFNN is very close to the reference. However, the lines based on $Acc_r$ obtained via LF data and without $Acc_r$ are markedly different from the line of reference, which indicates that these two cases are not accurate.

## 4. Conclusions

In this paper, we propose an accurate and computationally-feasible deep learning model, i.e., MFNN, for constructing a proppant settling model in one-dimensional and two-dimensional parameter space. MFNN combines the variation trends of LF data and the accuracy of HF data. The HF data and LF data are generated through unresolved CFD-DEM simulations. We construct and validate the particle settling model in one-dimensional parameter space and two-dimensional space through MFNN, respectively. In addition, we discuss the application of the particle settling model to macro-scale proppant simulations. According to the results, the following conclusions can be drawn:

(1) Although there is some error in the LF simulations, the results of the LF simulations are similar to those of the HF simulations in terms of variation trends, which has some reference value for prediction. The behavior of the particle cluster of LF simulations is also similar to that of HF simulations.

(2) MFNN can construct an accurate particle settling model and reduce computational cost by combining LF data and HF data. In one-dimensional and two-dimensional parameter space, MFNN could save 80% computational time compared to HF simulations. Overall, the higher is the parameter dimension, the more efficient is MFNN.

(3) In the proppant transport process, proppant settling plays a crucial role and has a significant influence on proppant distribution in inclined fractures, which cannot be neglected. Irrespective of the fracture angle, proppant settling could lead to over 80% relative error in terms of the height of the sand bed.

(4) The results of simulations using the proppant settling model constructed by MFNN are very close to the HF data. The relative error is less than 1%, while the application of LF data can lead to large relative error, i.e., even up to 50%.

In aggregate, MFNN opens new pathways for rapidly constructing particle settling models in inclined fractures. However, one potential challenge for the proposed method is that it may be difficult to validate the settling model in high-dimensional parameter space because the validation requires a large number of HF simulations. Moreover, the present research focuses on inclination angle and particle density ratio. Future work should indeed focus on more extensive parameter space. In addition, the sampling method utilized in this paper is random sampling, which would have a significant effect on the model performance and result in waste of computational resources. In the future, we should attempt to optimize the sampling method to identify the most valuable samples in order to improve the accuracy and efficiency of the model.

## Acknowledgments

This work is partially funded by the Shenzhen Key Laboratory of Natural Gas Hydrates (Grant No. ZDSYS20200421111201738) and the SUSTech - Qingdao New Energy Technology Research Institute.